# Controlling light in scattering media noninvasively using the photo-acoustic transmission-matrix

T. Chaigne†, O. Katz†, A.C. Boccara, M. Fink, E. Bossy, S. Gigan*

Institut Langevin, UMR7587 ESPCI ParisTech and CNRS, INSERM ERL U979 , 1 Rue Jussieu, 75005 PARIS

* sylvain.gigan@espci.fr

† These authors contributed equally to this work

**Optical wavefront-shaping has emerged as a powerful tool to manipulate light in strongly scattering media[1, 2]. It enables diffraction-limited focusing[3] and imaging[4, 5] at depths where conventional microscopy techniques fail[6]. However, while most wavefront-shaping works to-date exploited direct access to the target[2-5, 7-11] or implanted probes[12, 13], the challenge is to apply it non-invasively inside complex samples. Ultrasonic-tagging techniques have been recently demonstrated[14-18] but these require a sequential point-by-point acquisition, a major drawback for imaging applications. Here, we introduce a novel approach to non-invasively measure the optical transmission-matrix[5] inside a scattering medium, exploiting the photo-acoustic effect[19-23]. Our approach allows for the first time to simultaneously discriminate, localize, and selectively focus light on multiple targets inside a scattering sample, as well as to recover and exploit the scattering medium properties. Combining the powerful approach of the transmission-matrix with the advantages of photoacoustic imaging[19-21] opens the path towards deep-tissue imaging and light-delivery utilizing endogenous optical contrast.**

The microscopic-scale inhomogeneity of complex samples such as biological tissues leads to light scattering, which poses the main limitation on the penetration depth of current optical microscopy and laser nanosurgery techniques[6]. Scattering limits light focusing by diffusing any propagating beam, resulting in a decrease in intensity and loss of spatial resolution. On the micron scale, interferences between the scattered light components are manifested as random speckle patterns[24]. Although usually disregarded in deep-tissue diffused-light imaging techniques, these complex interferences are deterministic. They can thus be coherently manipulated and exploited for focusing and imaging by wavefront-shaping, using computer-controlled spatial light modulators (SLMs)[1] or phase-conjugate mirror[14, 25].

The concept of focusing scattered light by wavefront shaping is based on utilizing a feedback signal for the light intensity at the target point. Measuring this signal for different input wavefront patterns shaped by an SLM, allows one to maximize the intensity on the target either through iterative optimization[2-4, 8, 9], phase-conjugation[13-16] or by calculating the complex relation between input and output modes[5, 7, 10]. The latter, "optical transmission-matrix" approach[5], enables going much beyond single-target focusing, and allows to generate any desired light pattern on multiple targets, as well as to study the sample scattering properties, such as the optical memory-effect and transmission eigenchannels[26, 27]. In most wavefront-shaping demonstrations to date the feedback signal was provided by a camera placed at the target plane behind the scattering sample[2, 5, 8, 9]. Such geometry is not applicable in practical scenarios where the target plane is not directly accessible. To focus *inside* a scattering sample, one has to find an alternative method to monitor the optical intensity at the target points. Implanting fluorescent or second-harmonic "guide stars"[12, 13] is one successfully studied path, but in addition to being invasive, it only allows focusing in the vicinity of these static targets. Ultrasonic-tagging the target zone using acousto-optic techniques offers dynamic and flexible control over the probe position[14-18], but in addition to the resolution limit imposed by the acoustic wavelength, each acousto-optical measurement is limited to a single target. Thus, utilizing these techniques for imaging or laser nanosurgery where the investigation of a large number of target points is mandatory, involves very long acquisition times, where all points of interest have to be sequentially tagged and measured[14-18].

Here, we propose a novel approach that allows for the first time to *simultaneously* and non-invasively detect, localize, and selectively focus light on *multiple* targets in a large volume inside a scattering medium, as well as to study scattering properties of the medium as manifested inside the sample. We achieve this by exploiting the photoacoustic effect[19] to measure the transmission-matrix *inside* a sample, i.e. by measuring the acoustic waves that are generated by absorbing targets for different input illuminating wavefronts (Fig.1a). This method does not require direct optical access to the targets as the acoustic waves are remotely measured by an ultrasonic transducer placed outside the sample[19]. We show here that photoacoustics enables effective measurement of the transmission-matrix even when the spatial resolution of the acoustic detection is lower than the optical mode size (speckle grain size) in the medium. While photoacoustic feedback has been investigated for single-point focusing with an optimization-based approach[22], here, we demonstrate multiple-points selective focusing through scattering samples including 0.5mm-thick chicken breast tissue. In addition we retrieve the optical 'memory-effect'[28] directly from the measured photo-acoustic matrix, and use the singular value decomposition of the transmission matrix to discriminate and localize the absorbing targets[29].

## Results

**Principle**

To selectively focus light on any single target from a large number (*M*) of target points inside a scattering medium, one has to know the input-output phase relations $\varphi_{mn}$ between the electric fields of each input mode (SLM pixel) $E_n^{in}$ (*n=1..$N_{SLM}$*), and each target point (output mode) $E_m^{out}$ (*m=1..M*), where $N_{SLM}$ is the number of SLM pixels. Considering a linear propagation medium, this relationship is given by the complex-valued *M*×*N* optical transmission-matrix[5] *T* with elements $t_{mn} = |t_{mn}|e^{i\varphi_{mn}}$. The complex optical field at the *m*th output position is then given by $E_m^{out} = \sum_{n=1}^{N_{SLM}} t_{mn} E_n^{in}$.

In all works to date, measuring the transmission-matrix *T* has been realized by imaging the target plane on a high-resolution camera[5, 7, 10, 11]. *T* can be directly retrieved from the camera images by phase-shifting interferometry[5]. In this measurement process, the intensities of all output modes are monitored as a function of a phase modulation $\varphi_n^{SLM}$ on the *n*th input mode. Shifting the phase $\varphi_n^{SLM}$ from 0 to 2π will result in a cosine-modulation of the output modes intensities. The modulation amplitudes and phase-retardances provide $|t_{mn}|$ and $\varphi_{mn}$, respectively. Repeating this measurement step for *n=1* to *n=$N_{SLM}$* gives the required influence of the SLM pixels on *all* output modes. This is in contrast to iterative optimization approaches[2, 22] where the same number of measurements is required for *each* output mode. The main drawback of the transmission-matrix approach was, until today, the requirement to directly (and thus invasively) optically image the output plane with a camera. Nonetheless, here we demonstrate that the transmission-matrix measurement scheme can be applied noninvasively in practical scenarios by *replacing the camera with a time-resolved photoacoustic signal detection*.

In photoacoustics[19], when a laser pulse hits an absorbing portion of the sample (which can be for example a small blood vessel[20]), an ultrasound pulse is generated through transient thermo-elastic stress generation that propagates away from the absorber. The amplitude of the photo-acoustic wave varies linearly with the light intensity on the absorber. Because the acoustic pulse propagates in soft tissues with negligible scattering it can be measured non-invasively by an ultrasonic transducer placed outside the sample[19] (Fig.1a). After a single laser shot, the photo-acoustic waves from *all* absorbing targets in the transducer's field-of-view reach the transducer with time-of-arrivals $\tau_m$ corresponding to their distances $z_m$ from the transducer $z_m = c_S \tau_m$, where $c_S$ is the speed of sound in the medium (Fig.1b). Thus, the time-resolved signal trace from a single spherically focused acoustic transducer can serve, in essence, as a row of pixels of a virtual camera sensing the spatial light distribution inside the medium, with the ultrasound resolution. Interestingly, unlike acousto-optical tagging, the minimum size of the photoacoustic probed volume is not limited by the acoustic wavelength but rather by the absorbers dimensions in the case of sub-ultrasound-wavelength absorbers.

A time-resolved photoacoustic trace (Fig.1b) can replace the camera signal for the non-invasive measurement of the transmission-matrix even when extended targets are considered. The main difference being that the resolution of the acoustic probing may be larger than the optical mode size (speckle grain) in the medium. Thus, the photoacoustic signal amplitude at each time delay $V_{PA}(\tau_m)$ is proportional to the sum of *several* optical modes intensities contained in the corresponding probed volume, dictated by the ultrasound resolution. However, thanks to linearity, modulating each SLM input mode phase, as in the standard transmission-matrix measurement, still results in a cosine modulation of $V_{PA}(\tau_m)$ (Fig.1c). This is easily understood as $V_{PA}(\tau_m)$ corresponds to the linear sum of the cosine modulated acoustic emission of all absorbers contained in the probed volume (see detailed analysis in Supplementary Equations 1-6). As in the all-optical transmission-matrix approach[5], the cosine modulation amplitudes and phase-retardances of $V_{PA}(\tau_m)$ give the photo-acoustic transmission-matrix coefficients for the probed volume corresponding to the output position around $z_m = c_S \tau_m$. The probed volume dimensions are limited transversely ($\Delta x$) by the transducer acoustic frequency $f$ and numerical aperture: $\Delta x = (c_S/f) \cdot (F/D)$, and axially ($\Delta z$) by the transducer's impulse response: $\Delta z = c_S \Delta \tau \approx c_S/(f \times B)$, where $F/D = 2$ is the transducer's F-number, $f$ is its central frequency and B its relative bandwidth (typically on the order of 100% for broadband imaging transducer).

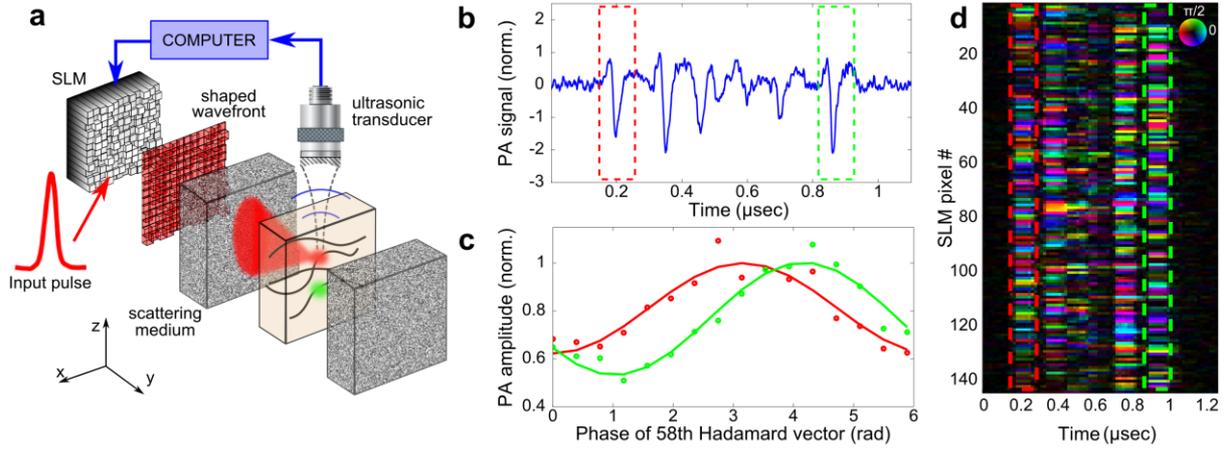

**Figure 1 : Measuring the transmission matrix photo-acoustically. a,** A pulsed laser beam is shaped by an SLM and illuminates an agarose gel phantom placed behind a scattering sample. A spherically focused ultrasonic transducer measures the photoacoustic signals generated by absorbing 30μm-diameter wires embedded in the sample; **b,** temporal trace of the experimentally measured photoacoustic signal following a single laser shot through a scattering diffuser; the signals at the different time delays correspond to different wires located at different distances from the transducer; **c,** The evolution of the peak-to-peak signal amplitudes for two wires, corresponding to the two framed time-windows shown in b, while scanning the phase of a single input mode on the SLM (in the Hadamard basis) from 0 to $2\pi$. The signal from each target closely follows a cosine modulation as a function of each input mode phase (Supplementary Eq.2); **d,** The photoacoustic transmission-matrix is retrieved directly from the measured modulation phases and amplitudes of (c) for all input modes. The transmission-matrix describes the influence of each input mode (in the SLM pixel basis, vertical axis) on each acoustic voxel (time window in photoacoustic trace, horizontal axis). The same framed time windows of **b**, are framed.

## Experimental transmission-matrix measurement

To experimentally demonstrate the noninvasive photo-acoustic transmission-matrix measurement approach, an agarose gel phantom containing 30μm-diameter absorbing nylon wires was illuminated through a scattering sample by a 5ns pulsed laser source (Fig.1a, see Methods for details). The laser beam was wavefront shaped by a phase-only SLM. A spherically-focused ultrasonic transducer (30MHz central frequency, 30MHz -6dB bandwidth, F-number=2) detected the time-dependent photoacoustic signals generated by the wires, which could for instance mimic capillary vessels, a target used to study angiogenesis with photoacoustic imaging[20, 21]. The recorded trace following each laser pulse contained several signals generated by multiple targets with the time delays corresponding to their positions along the axis of the transducer (Fig.1b). To measure the transmission-matrix, the procedure presented by Popoff et al.[5] was employed. In this measurement procedure the phase of each of the $n=1$ to $n=N_{SLM}=140$ input modes was sequentially modulated from 0 to $2\pi$ in ≥3 steps. For each input-mode phase $\varphi_n^{SLM}$, a photoacoustic signal trace $V_{PA,n}(\tau)$ was recorded (Fig.1b), digitally filtered and the peak-to-peak signal amplitude at each 90ns time-window was computed. This time-window width was carefully chosen to match the typical impulse response of the transducer for a single isolated absorber, maximizing the signal-to-noise (see Supplementary Fig.2). To achieve the maximal modulation depth the Hadamard basis rather than the SLM-pixel basis was used as the input-mode basis (see Methods)[5].

As expected, the peak-to-peak signal at each of the time-windows followed a cosine modulation as a function of the input-mode phase (Fig.1c), confirming the linearity assumption in the theoretical analysis (Supplementary Eqs.1-3). The photoacoustic transmission-matrix $T^{PA}$ was directly obtained for all time windows (thus all absorbing targets) simultaneously from these modulation amplitudes and phases (Fig.1c-d). An example for the photo-acoustic transmission matrix obtained experimentally through an optical diffuser is presented in Fig.1d. This matrix describes the influence of each input mode (SLM pixel, vertical axis) on each of the absorbing targets along the transducer focus (time delay in photoacoustic trace, horizontal axis). A visual inspection of the measured matrix clearly reveals that the information on at least five individual absorbing targets is simultaneously retrieved at this specific transducer position.

**Controlled focusing using the transmission-matrix**

As a first demonstration for utilizing the measured photo-acoustic transmission-matrix we present deterministic selective light focusing on several absorbers along the focus of the ultrasonic transducer (Fig.2). The results for focusing obtained through an optical diffuser are presented in Fig. 2a-c, and the results obtained through a sample of ~500μm thick chicken breast tissue are presented in Fig.2d-f. Figures 2a,d show the experimentally measured transmission-matrices in the two experiments. The SLM phase-pattern required to selectively focus on a target located at a specific axial position $z_m=c_S\tau_m$ is simply the phase-conjugate of the transmission-matrix column corresponding to time delay $\tau_m$ (Figs.2b,e, Supplementary Eq.6)[5]. The photoacoustic signals obtained when displaying two of these focusing phase-patterns on the SLM are presented in Figs.2c,f. These confirm a selective enhancement of the photoacoustic signals of the selected targets.

The expected intensity enhancement for focusing with the all-optical transmission-matrix is given by[26]: $\eta=0.5N_{SLM}/N_{modes}$. It is proportional to the number of controlled degrees of freedom $N_{SLM}$, and inversely proportional to the number of optical modes (speckle grains) whose intensity is enhanced, $N_{modes}$. In our experiments $N_{modes}$ corresponds to the number of modes enclosed within the focusing resolution, $\Delta x \approx 100 \mu m$, i.e. the number of speckles contained in the target absorbing area intersecting the acoustic focus. In the results presented in Fig.2, $N_{modes} \approx 6$, and the expected intensity enhancement is $\eta \approx 0.5 \cdot 140/6 \approx 11.5$ (see Supplementary chapter 3), which is close to the experimental enhancement factors of $\eta \approx 6$. We note that with the knowledge of the TM, one is not limited to focusing on a single target and any intensity distribution on the targets can be generated[5, 7].

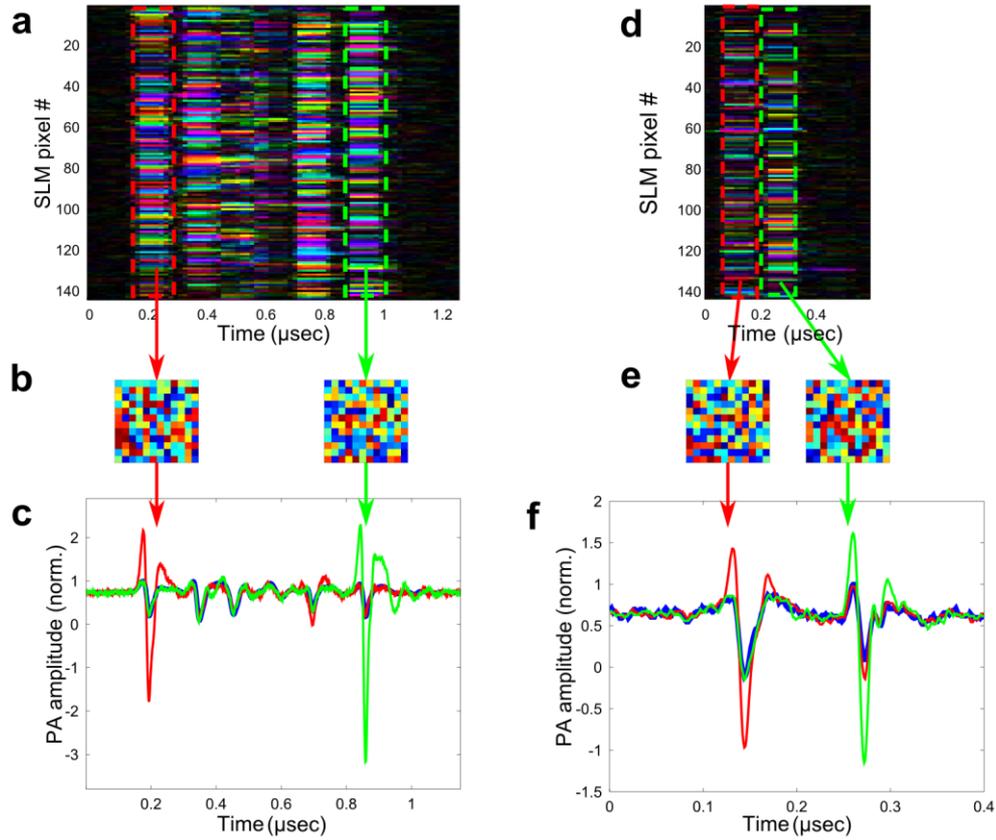

**Figure 2. Selective focusing with the transmission-matrix. (a-c)** Measurements through an optical diffuser: **a,** The measured transmission-matrix. Each entry gives the amplitude and phase of the modulation induced by each SLM pixel (vertical axis) on each time window in the photoacoustic signal (horizontal axis). **b,** The phase masks for focusing on the nearest (red) and farthest (green) absorbers are given by phase conjugating the corresponding transmission-matrix columns. **c,** The measured photoacoustic signals when displaying a flat phase SLM (blue), the phase-pattern of (b) for focusing on the first absorber (red), and the phase pattern of (b) for focusing on the farthest absorber (green) . **(d-f)** same as (a-c) but with a 0.5mm thick chicken breast tissue as the scattering sample; only two absorbing wires are in the focus of the transducer in this experiment.

## Retrieving scattering properties from the transmission-matrix: the optical 'memory-effect'

The information encoded in the optical transmission-matrix is much richer than the one that is required to focus on a single or multiple targets. Here we experimentally demonstrate that it allows the probing of the sample's optical memory-effect[28].

The 'memory effect' for speckle correlations is, in a nutshell, the fact that a multiply scattering sample of thickness L retains a range of isoplanatism, i.e., the speckle patterns generated by plane-waves at different incident angles $\theta_1$, $\theta_2$ will be correlated as long as $\Delta\theta = \theta_1 - \theta_2$ is smaller than $\sim \lambda/2\pi L$, where $\lambda$ is the wavelength of the incident light[28]. Similarly, in weakly aberrating samples this range is known as the isoplanatic-patch size both in optics[30] and acoustics[23]. The memory-effect is important not just because it is a mesoscopic property of the medium[31], but as it can serve to scan the focus obtained by wavefront-shaping[4, 13] and allow real-time imaging within the memory-effect range[9].

In the measured photo-acoustic transmission matrix as presented in Fig.1d, the phases of each column $m$, represents the SLM phase-pattern $\varphi_{SLM}^m(x,y)$ required to focus on a target absorber at an axial position $z_m=c_S\tau_m$ from the ultrasonic transducer (Figs.2b,3a). If these targets are within the sample's memory-effect range, the difference between each such two SLM phase-patterns would typically be a tilt (Fig.3b). The k-space representation of this tilt (obtained by spatial Fourier-transform of the complex exponentials of the phase-pattern) would then be a shifted delta-function.

Thus, to examine the memory-effect range one can plot the k-space representation of the phase-differences between the transmission-matrix columns (Fig.3c). In the case of perfect memory-effect speckle correlations, this representation would result in a matrix with non-zero values aligned on a diagonal, representing the angular (i.e. k-space) correlations between the scattered waves. We have performed this analysis on the experimental transmission-matrix measured through a thin diffuser (Fig.1d), for a tilt along the z direction, and obtained the results presented in Fig.3c. The result clearly reveals the k-space correlations of the diffuser's transmission-matrix columns, highlighting the large memory-effect range expected from such thin scatterer. The apparent wrapping of the large k-space components (k-space aliasing, or equivalently grating lobes) is a result of the low SLM-resolution used in this experiment (12x12 pixels). We note that the presented analysis enables for the first time to directly access the memory-effect angular range inside scattering samples, and is analogous to inspecting the speckle field-field correlations (See details in Supplementary Eqs.7-9).

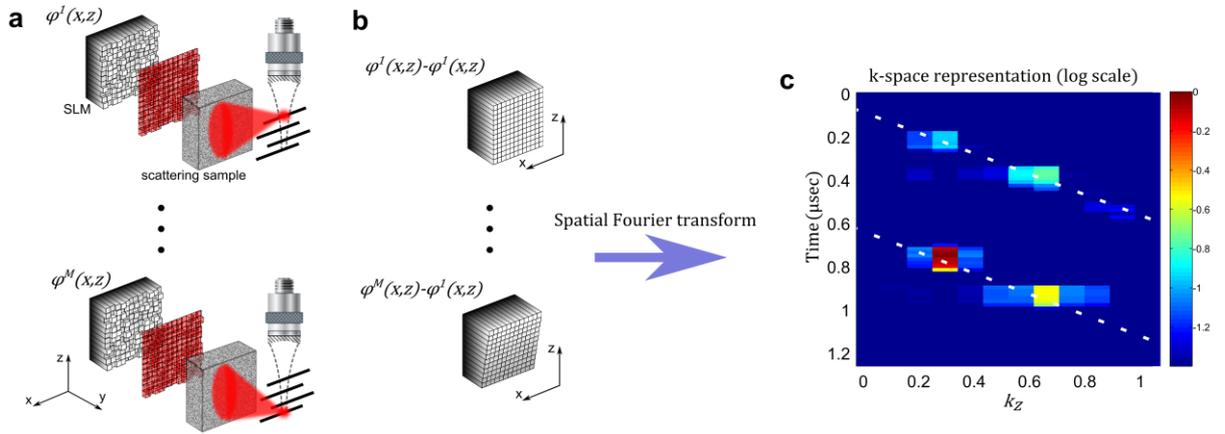

**Figure 3: Revealing the 'memory-effect' from the measured transmission-matrix.** A simple analysis of the measured transmission-matrix can reveal the existence of the angular memory-effect for speckle-correlations[28], expected in thin scattering samples. In this analysis, the phase difference between the transmission matrix columns (as presented in Fig.1d) is analyzed in $k_z$-space, corresponding to a tilt in the z-direction. **a,** Each transmission-matrix column $m$ is the SLM phase-pattern $\varphi^m(x,y)$ required to focus on the target located at a distance $z_m=c_S\tau_m$ from the transducer; **b,** The difference between two phase patterns required to focus on two targets which are within the memory-effect range is simply a tilt, i.e. a delta-function in k-space representation; the memory-effect is revealed by displaying this k-space difference for all the matrix columns. Within the memory effect range this representation results in a sparse matrix with non-zero entries on the diagonal. **c,** the result of the above analysis on the experimental transmission-matrix measured through a diffuser (Fig.1d), revealing the k-space correlations of the memory-effect between the different targets, expected for such a thin scatterer; dashed lines represent the diagonal direction for a perfect memory effect and its k-space aliasing (see text).

**Targets localization by singular value decomposition**

An extremely powerful tool in the analysis of the scattering matrix is the singular value decomposition (SVD). Recently, SVD of the optical transmission matrix was used to identify transmission-eigenchannels[5] and maximize energy transport in multiply-scattering samples[27]. Interestingly, the SVD of the backscattering-matrix was utilized to discriminate and selectively focus on individual nanoparticles through aberrating media[29], a result analogous to works done in ultrasound[32, 33]. Here we show that similarly, the SVD of the photo-acoustic transmission-matrix enables the identification and discrimination of individual absorbing targets without any a-priori information on their number or positions, giving the wavefronts required to selectively focus on these targets.

To achieve this, the SVD of the photoacoustic transmission-matrix measured through a diffuser (Fig.1d) was computed and analyzed (Fig.4). An SVD of a matrix $T$ consists of writing $T = U\Sigma V^*$, where $\Sigma$ is a rectangular diagonal matrix containing the real positive singular values, $\lambda_i$, in descending order, and $U$ and $V$ are unitary matrices whose columns corresponds to the output and input singular vectors, $U_i$ and $V_i$, respectively. Each input singular vector $V_i$ corresponds to the SLM phase-pattern for focusing on a target having the photoacoustic modulation intensity $\lambda_i$. The corresponding output singular vector $U_i$ is the expected photoacoustic modulation time-trace for this singular value, i.e. the corresponding absorber's position.

The results of this analysis are presented in Fig.4. Plotting the obtained singular values in descending order (Fig.4b), one can identify that their distribution exhibits two parts. A continuum of low singular values associated with background noise, and a few higher singular values, which are associated with strong absorbing targets (highlighted in Fig.4b). Plotting the output singular vectors, $U_i$, which correspond to these singular values, clearly reveals strong and temporally localized photoacoustic modulation for the singular values above the background noise (Fig.4a). Comparing the temporal positions of these modulations with the raw recorded photoacoustic signal trace (Fig.4c) shows an excellent correspondence between these singular vectors and the time delays where absorbing targets can be subjectively identified by visual inspection of the temporal trace. This proves that the SVD can be used to accurately give the number and position of the absorbing targets, within the limitation of the ultrasound resolution, i.e. one absorbing target per ultrasound resolution cell. Such an approach advantageously replaces the subjective imprecise visual inspection of the transmission-matrix used to select the time-delays for focusing in Fig.2. The corresponding *input* singular vectors $V_i$ can thus be used to 'automatically' guide light to these targets. Furthermore, because of the linearity of the problem, one can also use a combination of two or more singular vectors to focus light simultaneously on several absorbers.

We note that these results are analogous to the well-studied techniques in acoustics where the SVD of the time-reversal operator is used for focusing in a multi-target medium[28, 29], and to the recent implementation in optics[29]. The underlying enabling principle is that a one-to-one association exists between each eigenstate and a scattering target, and that in first approximation and when there is no degeneracy, each singular vector $V_i$ corresponds to the complex amplitude-mask which, when applied to the SLM, focuses onto the associated target.

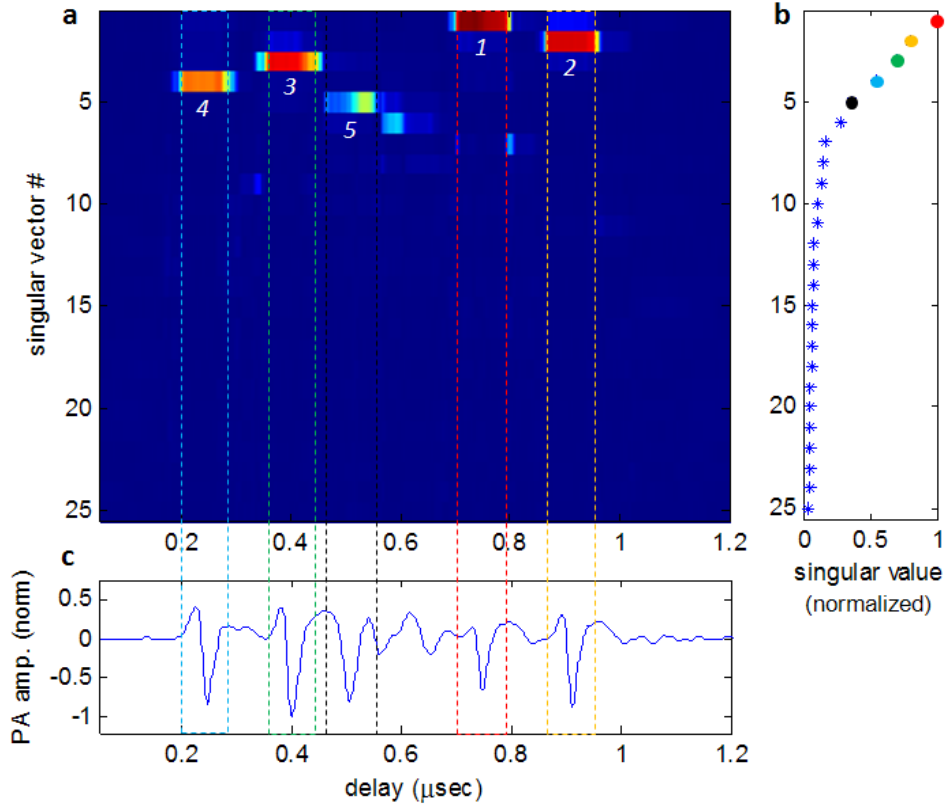

**Figure 4: Localizing absorbing targets by singular value decomposition (SVD) of the transmission matrix.** Results for the SVD of the photo-acoustic matrix measured through a diffuser (Fig.1d); **a,** The first 25 output singular vectors, $U_i$ weighted by their corresponding singular values, $\lambda_i$; **b,** the first 25 singular values, $\lambda_i$; A few singular values rise above the continuum of background-noise singular values; **c,** the raw average photoacoustic time-trace, showing an excellent correspondence between the output singular vectors of (a) and the time delays where absorbing targets can be subjectively identified.

# Discussion

In this work we presented a novel non-invasive approach for controlling light inside a scattering medium. This method enables for the first time controlled focusing on multiple targets and the study of scattering media using the photo-acoustic transmission matrix. Although not demonstrated in this work, the obtainable optical focus resolution is limited only by the size of the probed absorbers and therefore can inherently surpass the ultrasound-diffraction limit.

While the proof-of-principle was demonstrated here for focusing through scattering samples, the proposed approach equivalently applies to focusing inside scattering media, as the photo-acoustic feedback is obtained remotely with no direct access to the target. In order to apply the technique to focus with high efficiency inside strongly scattering media such as biological tissues, one has to maintain a high ratio between the number of controlled degrees of freedom to the number of optical modes on each target within the acoustic focus. Given the sub-micron diffraction-limited dimensions of optical modes inside tissues, this goal can be met by the combined use of a high resolution SLM, absorbing targets of small dimensions and a high-frequency acoustic detection.

Another practical aspect is the reduction of the transmission-matrix acquisition time, which would be valuable for real-time or in-vivo applications. In this work, the acquisition time was limited by the repetition rate of the laser used (10Hz). The use of lasers with higher repetition rates in combination with fast SLMs[34] is expected to significantly reduce the acquisition times. Because the total acquisition time is dictated by the number of measurements used per controlled DOF, a convenient trade-off exist between the focusing-efficiency and the acquisition time, which can be particularly advantageous in certain applications.

We used photoacoustic feedback to perform light focusing at depth, but we envisage that the method demonstrated here would also benefit the field of photoacoustic imaging: while photoacoustic imaging is now a mature technique for tissue imaging at large optical depths[21], it currently relies on diffuse homogeneous illumination. Our method opens up the possibility of spatially optimizing the light delivery to the regions of interest, hence opening the possibility to reveal different photoacoustic features at even larger depths compared to current photoacoustic approaches.

Finally, we note that the advantage of the presented approach is in its generality, as it only requires access to relative variations of a linear optically-induced signal from a given location. In particular, while demonstrated here with a single element focused ultrasonic transducer, it can be directly implemented on a photoacoustic imaging setup using an array of ultrasound transducers[23]. One can also imagine extending the matrix method to other contrast mechanisms that can be probed locally and noninvasively, e.g. the real-time probing of the local heat deposition via magnetic resonance imaging[35].

## Methods

**Experimental setup.** The complete experimental setup is described in Supplementary Fig. 1. An agarose gel phantom containing 30μm-diameter absorbing nylon wires (NYL02DS, VetSuture) was placed behind the scattering sample. The scattering samples were 0.5° Newport light shaping diffuser, which spread the light evenly with no notable ballistic component, and a ~0.5mm thick chicken breast sample, sandwiched and partially dried between two glass slides. The phantom was illuminated through the scattering sample by an attenuated beam from a 5ns pulsed laser source (Continuum Surelight, 10 Hz repetition rate, <10mJ pulse energy at a wavelength of 532nm), shaped by a phase-only SLM (Boston Micromachines Multi-SLM). The photoacoustic signals generated by the absorbing wires were detected by a focused ultrasonic transducer (SNX110509_HFM13, Sonaxis), having a central frequency of 30MHz and equivalent bandwidth, with an F-number of 2 and 7.5mm focal length. The signal from the transducer was amplified through a preamplifier (Model 5900PR, Sofranel) and recorded on an oscilloscope (Lecroy WaveSurfer 104 MXs-B). The oscilloscope signal was sampled at 1GHz, and digitally filtered by a bandpass filter between 2 and 60MHZ.

**Transmission matrix measurement.** The phases of each Hadamard input vectors were ramped in 16 steps between 0 and 2π, and for each step the photo-acoustic signals corresponding to 5 successive laser pulses were recorded and averaged. The modulation amplitude and phase for each of the Hadamard input vector and output temporal window was extracted using a discrete Fourier transform (Fig.1c, see Supplementary Eq.4). Once the transmission-matrix was measured in the Hadamard input basis, it was converted to the canonical "input SLM-pixels" basis by a unitary transformation (Hadamard-canonical basis transformation, as was done in ref[5]).

**Acknowledgements**

We thank J. Gateau for fruitful discussions and for valuable comments on the manuscript, and David Martina for technical support. This work was funded by the European Research Council (grant number 278025) and by the Fondation Pierre-Gilles de Gennes pour la Recherche.

**Author contribution**

S.G. and E.B. designed the initial experimental setup. O.K. proposed the photoacoustic transmission matrix approach and analysis. S.G., E.B, T.C and O.K. discussed the experimental implementation, T.C. and O.K. performed the experiments and analyzed the results. All authors contributed to discussing the results and writing the manuscript.

# Controlling light in scattering media noninvasively using the photo-acoustic transmission-matrix – Supplementary information

T. Chaigne†, O. Katz†, A.C. Boccara, M. Fink, E. Bossy, S. Gigan

Institut Langevin, UMR7587 ESPCI ParisTech and CNRS, INSERM ERL U979, 1 Rue Jussieu, 75005 PARIS

## 1. Experimental setup

A detailed sketch of the experimental setup is presented in Supplementary Figure 1. The beam from the nanosecond Q-switched laser (Continuum Surelight, 10 Hz repetition rate, 5ns pulses at 532nm wavelength), was magnified by a ×5 telescope (composed of lenses L1, L2). The pulse energy was attenuated to <10mJ using a half-wave plate and a polarizing beam-splitter (PBS). The beam was size was then set using an iris to illuminate the full aperture of the SLM (Boston Micromachines Multi-SLM). The SLM surface was imaged on the surface of the scattering sample using a 4-f telescope composed of lenses L3 and L4, with a demagnification of ~13. The 4-f imaging configuration was used to ensure a constant illumination, fixed optical speckle size under different applied SLM phase-patterns, and a maximum range of the optical memory-effect[1]. In the experiments with the optical diffuser, an additional focusing lens was placed between the SLM and L3 to add a spherical curvature on the input beam, assuring a strong contribution of all the controlled degrees of freedom on the target.

A small portion of the beam was directed toward a photodiode which monitored each pulse intensity variation and was used to normalize the measured photoacoustic signals against pulse-to-pulse intensity variations. The photoacoustic signal was measured by a focused ultrasonic transducer (SNX110509_HFM13, Sonaxis), and amplified using a Sofranel Model 5900PR preamplifier. Both the photodiode and photoacoustic signals following each laser pulse were digitized by a Lecroy WaveSurfer 104 MXs-B oscilloscope and saved in real-time by a personal computer.

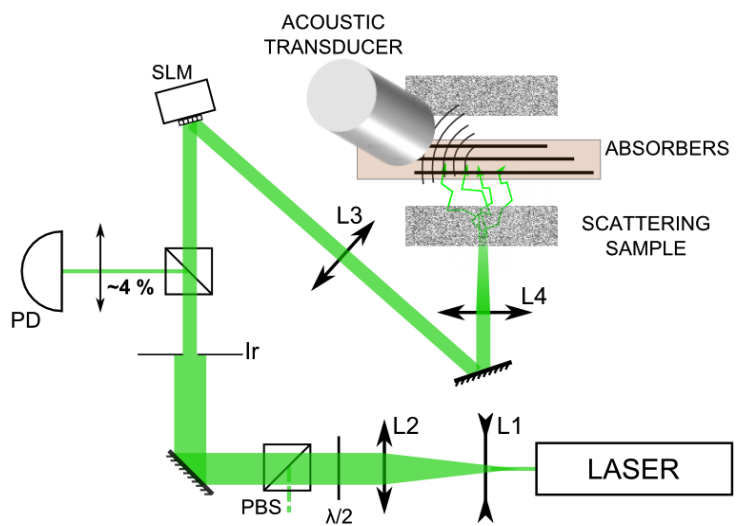

**Supplementary Figure 1: Experimental Setup:** L1: f=-50mm, L2: f=250mm, λ/2: half-wave plate, PBS: polarizing beam splitter, Ir.: iris, SLM: spatial light modulator, L3: f=300mm, L4: Bausch & Lomb objective, 8x, 0.15 NA.

## 2. Transmission matrix measurements

As explained in the manuscript main text, the photoacoustic transmission-matrix is obtained by measuring the modulation of the photoacoustic signal when the conventional phase-shifting method for measuring the transmission-matrix is employed[2], i.e. when several phase-shifts of 0 to $2\pi$ are applied to each input Hadamard vector on the SLM. In this section we analyze the theoretical expected modulation in the measured photoacoustic signal using this scheme and detail how the photoacoustic transmission-matrix coefficients were obtained from the measured signal.

As mentioned in the main text[2] the optical field at the $m^{th}$ output position inside the medium is given by the *optical* transmission-matrix, $t_{mn}$: $E_m^{out} = \sum_{n=1}^{N_{SLM}} t_{mn} E_n^{in}$, where $E_n^{in}$ is the light field on the $n^{th}$ pixel of the SLM. The measured photoacoustic signal amplitude at any time delay $\tau_m$ following the laser pulse is, on first approximation, proportional to the linear sum of all the optical modes intensities contained within the absorbing area in the probed acoustic volume $\nu \approx \Delta z \cdot \frac{\pi}{4} \Delta r^2 \approx (c_S \cdot \Delta \tau) \cdot \frac{\pi}{4} \left(\frac{c_S F}{fD}\right)^2$, centered around the axial position $z_m = c_S \tau_m$ from the transducer, where $c_s$ is the speed of sound in water, $f$ is the transducer's central frequency, F is the transducer's focal length, $D$ is the transducer's diameter, and $\Delta \tau = 1/Bf$ is the bandwidth limited temporal resolution of the transducer, according to the impulse response of the system. Thus, the *measured photoacoustic signal amplitude* at a time delay $\tau_m$, is proportional to the a linear sum of the optical intensities over all the modes impinging on absorbing targets contained in the corresponding volume $\nu(\tau_m)$, centered around $z_m = c_S \tau_m$:

$$V_{PA}(\nu(\tau_m)) \propto \sum_{m' \in V(\tau_m)} |E_{m'}^{out}|^2 = \sum_{m' \in V(\tau_m)} \left| E_{m'}^{ref} + \sum_{n=1}^{N} t_{m'n} E_n^{in} \right|^2 \quad (1)$$

where $E_{m'}^{out}$ is the light amplitude on each $m'$ of these absorbers and $E_{ref}$ is a constant background reference light field, originating from the unshaped part of the input field[2].

When the phase-shifting transmission-matrix measurement protocol is performed, each component of the input field is sequentially phase shifted from 0 to $2\pi$ compared to the reference part of the beam ($E_{n'}^{in} \propto \delta_{n,n'} e^{i\varphi_n^{SLM}}$). As a result, during this process the measured photoacoustic signal amplitude at any time delay $\tau_m$ evolves according to:

$$V_{PA,n}(\nu(\tau_m)) \propto \sum_{m' \in V(\tau_m)} \left| E_{m'}^{ref} + t_{m'n} e^{i\varphi_n^{SLM}} \right|^2 = \sum_{m' \in V(\tau_m)} \{a_{m'n} + b_{m'n} \cos(\varphi_{m'n} + \varphi_n^{SLM})\} \quad (2)$$

with $a_{m'n} = \left|E_{m'}^{ref}\right|^2 + |t_{m'n}|^2$, and $b_{m'n} = 2\left|E_{m'}^{ref} t_{m'n}\right|$.

It then follows:

$$V_{PA,n}(\nu(\tau_m)) \propto \alpha_{mn} + \beta_{mn} \cos(\theta_{mn} + \varphi_n^{SLM}) \quad (3)$$

where $\alpha_{mn} = \sum_{m' \in V(\tau_m)} a_{m'n}$,
$\beta_{mn} \cos(\theta_{mn}) = \sum_{m' \in V(\tau_m)} b_{m'n} \cos(\varphi_{m'n})$,
and $\beta_{mn} \sin(\theta_{mn}) = \sum_{m' \in V(\tau_m)} b_{m'n} \sin(\varphi_{m'n})$.

Thus the obtained result of Supplementary Eq.3 is that the measured photoacoustic signal, although being the sum of many modes, still follows a cosine modulation as a function of the input phase shift $\varphi_n^{SLM}$. The main practical difference between this photoacoustic modulation and the modulation in the all-optical transmission-matrix experiment being that the modulation depth (contrast), $\frac{\beta_{mn}}{\alpha_{mn}}$ is reduced compared to the direct measurement of the optical transmission-matrix.

In our experiments $\varphi_n^{SLM}$ was ramped between 0 and $2\pi$ in 16 equally spaced steps, giving a signal $V_{PA}^k(v(\tau_m))$ corresponding to $\varphi_n^{SLM} = \frac{2\pi k}{16}$. The photoacoustic transmission-matrix coefficients are defined as $t_{mn}^{PA} = \beta_{mn} e^{i\theta_{mn}}$. We have retrieved these by performing a discrete Fourier transform after completing each input mode 0 to $2\pi$ phase modulation (analogous to a lock-in detection):

$$\theta_{mn} = \arg\{t_{mn}^{PA}\} = -\arg\left\{\sum_{k=1}^{16} V_{PA,n}^k(v(\tau_m)) \times e^{i\frac{2\pi k}{16}}\right\} \quad (4)$$

$$\beta_{mn} = |t_{mn}^{PA}| = \left|\sum_{k=1}^{16} V_{PA,n}^k(v(\tau_m)) \times e^{i\frac{2\pi k}{16}}\right| \quad (5)$$

$t_{mn}^{PA}$ contains all of the information required to selectively focus light onto any of the targets within the focal zone of the acoustic transducer:

$$\varphi_n^{SLM, focusing\ at\ target\ m} = -\theta_{mn} = -arg\{t_{mn}^{PA}\} \quad (6)$$

In our experiments, to maximize the signal to noise, we have chosen to retrieve the transmission-matrix not for every time delay $\tau_m$ separately, but instead chose the photoacoustic signal to be used as the peak-to-peak amplitude in each 90ns time-window over the measured photoacoustic time trace with 15ns steps. The 90ns time-window was selected to match the photoacoustic impulse response of a single absorber as measured by the transducer in a calibration experiment (Supplementary Fig.2).

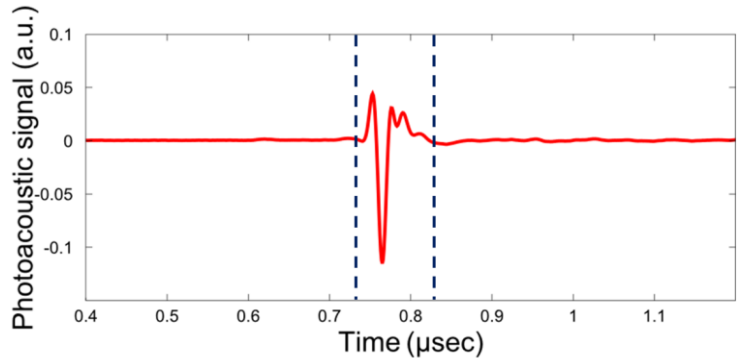

**Supplementary Figure 2: Measured photoacoustic impulse response for a single isolated absorbing target:** typical acoustic pulse from a single absorber. The framed part corresponds to the 90ns window scanned over the whole time trace to measure the transmission matrix.

## 3. Focusing intensity enhancement estimation

To estimate the expected photoacoustic signal enhancement, $\eta=0.5N_{SLM}/N_{modes}$[3, 4], one has to calculate the number of independent optical modes contained within the absorbing area of the target inside the acoustic focus. The absorbing target used are wires with a diameter of $\emptyset_{absorber} = 30\mu m$. The diameter of the acoustic focus is given by $\emptyset_{acoustic\ focus} = \frac{c_s F}{fD} = \frac{2\times 1480\ m.s^{-1}}{30\cdot 10^6 Hz} = 100\mu m$, where F/D=2 is the F-number of the acoustic transducer and $f$ is its central frequency. Thus, in the case of a straight wire crossing the center of the acoustic focus, the absorbing area measured is 3000μm².

The measured speckle size at the target plane in the experimental geometry presented in Supplementary Figure 1 was $\emptyset_{speckle} \approx 25\mu m$ (FWHM/$\sqrt{2}$ of the intensity autocorrelation function, as measured by a CCD). Thus, the number of speckle grains contained in the effective absorbing region (i.e. the portion of the absorber that is listened to by the transducer) is $N_{modes} \approx \frac{3000\mu m^2}{\pi(12.5\mu m)^2} \approx 6$.

The expected theoretical enhancement factor is then $\eta = \frac{0.5 N_{SLM}}{N_{modes}} \approx \frac{0.5 \cdot 140}{6} \approx 11.5$, which is close to the obtained experimental factor $\eta_{exp} = \frac{A_{final}^{PA}}{A_{initial}^{PA}} \approx 6$.

## 4. Analysis for revealing the memory effect k-space correlations

In the measured photoacoustic transmission matrix, as plotted in Fig.1d, each column, i, represents the SLM phase-pattern $\varphi_i^{SLM}(x^{in}, z^{in})$ required to focus on a target at an axial distance $z_i^{out} = c_s \tau_i$ from the transducer. If the different probed targets are within the sample's memory-effect range, the difference between each such two SLM phase-patterns would be just the addition of a tilt, hence a linear phase-ramp (Fig.4a):

$$\varphi_i^{SLM}(x^{in}, z^{in}) - \varphi_j^{SLM}(x^{in}, z^{in}) = k_{ij,x} x^{in} + k_{ij,z} z^{in} + \varphi_0 \qquad (7)$$

Where **k**$_{ij}$=(k$_{ij,x}$,k$_{ij,z}$) is the wave-vector defining the additional required tilt.

Thus, in the case of perfect correlations, the k-space representation of this difference between the input wavefronts (transmission-matrix columns) would be a delta-function at the relative tilt:

$$FT\left\{e^{i\varphi_i^{SLM}(x^{in},z^{in})} e^{-i\varphi_j^{SLM}(x^{in},z^{in})}\right\} \propto \delta(\boldsymbol{k} - \boldsymbol{k}_{ij}) \qquad (8)$$

Plotting the transmission matrix columns after subtraction of the first column phase, would thus result in a diagonal transmission matrix, if the output basis is in the k-space as well. As the output basis in the matrix presented in Fig.3c is the photoacoustic time-delay, which corresponds to vertical position, we are able to probe the memory effect only for a vertical tilt of the input beam, i.e. in k$_z$ representation. The diagonal nature of the transmission-matrix is characteristic of the memory-effect, and it denotes the fact that the scattering medium behaves the same for the different incident angles.

The information obtained through this analysis is equivalent to analyzing the electric field correlations between the fields propagating from these targets through the medium. This can be seen directly from Eq.8 using the convolution theorem:

$$FT\left\{e^{i\varphi_i^{SLM}(x,z)} e^{-i\varphi_j^{SLM}(x,z)}\right\} = FT\{E_i(x,z) E_j^*(x,z)\} = corr\{\widetilde{E}_i(k_x, k_z), \widetilde{E}_j(k_x, k_z)\} \qquad (9)$$